\documentclass{blois}

\DeclareUnicodeCharacter{2212}{-}
\DeclareUnicodeCharacter{03A3}{-}
\DeclareUnicodeCharacter{039B}{-}

\def\be{\begin{equation}}
\def\ee{\end{equation}}
\def\bea{\begin{eqnarray}}
\def\eea{\end{eqnarray}}

\usepackage{amsmath}
\usepackage{ifthen}
\newboolean{uprightparticles}
\setboolean{uprightparticles}{false}
\usepackage{xspace} 
\usepackage{upgreek}

\def\lhcb   {\mbox{LHCb}\xspace}

\def\lhc    {\mbox{LHC}\xspace}

\def\MagUp {\mbox{\em Mag\kern -0.05em Up}\xspace}

\ifthenelse{\boolean{uprightparticles}}%
{

 \def\Ppi         {\ensuremath{\uppi}\xspace}

 \def\PDelta      {\ensuremath{\Delta}\xspace}                 
 \def\PXi         {\ensuremath{\Xi}\xspace}                 
 \def\PLambda     {\ensuremath{\Lambda}\xspace}                 
 \def\PSigma      {\ensuremath{\Sigma}\xspace}                 
 \def\POmega      {\ensuremath{\Omega}\xspace}                 
 \def\PUpsilon    {\ensuremath{\Upsilon}\xspace}

 \def\PB      {\ensuremath{\mathrm{B}}\xspace}                 
                  
 \def\PD      {\ensuremath{\mathrm{D}}\xspace}

 \def\PK      {\ensuremath{\mathrm{K}}\xspace}

 \def\Pb      {\ensuremath{\mathrm{b}}\xspace}                 
 \def\Pc      {\ensuremath{\mathrm{c}}\xspace}

 \def\Pi      {\ensuremath{\mathrm{i}}\xspace}

 \def\Pp      {\ensuremath{\mathrm{p}}\xspace}

 \def\Ps      {\ensuremath{\mathrm{s}}\xspace}                 
                  
 \def\Pu      {\ensuremath{\mathrm{u}}\xspace}

 \def\thebaroffset{0.0em}
}
{

 \def\Ppi         {\ensuremath{\pi}\xspace}

 \mathchardef\PDelta="7101
 \mathchardef\PXi="7104
 \mathchardef\PLambda="7103
 \mathchardef\PSigma="7106
 \mathchardef\POmega="710A
 \mathchardef\PUpsilon="7107
                  
 \def\PB      {\ensuremath{B}\xspace}                 
                  
 \def\PD      {\ensuremath{D}\xspace}

 \def\PK      {\ensuremath{K}\xspace}

 \def\Pb      {\ensuremath{b}\xspace}                 
 \def\Pc      {\ensuremath{c}\xspace}

 \def\Pi      {\ensuremath{i}\xspace}

 \def\Pp      {\ensuremath{p}\xspace}

 \def\Ps      {\ensuremath{s}\xspace}                 
                  
 \def\Pu      {\ensuremath{u}\xspace}

 \def\thebaroffset{0.18em}
}
\newcommand{\offsetoverline}[2][\thebaroffset]{\kern #1\overline{\kern -#1 #2}}%

\makeatletter
\ifcase \@ptsize \relax
  \newcommand{\miniscule}{\@setfontsize\miniscule{4}{5}}
\or
  \newcommand{\miniscule}{\@setfontsize\miniscule{5}{6}}
\or
  \newcommand{\miniscule}{\@setfontsize\miniscule{5}{6}}
\fi
\makeatother

\DeclareRobustCommand{\optbar}[1]{\shortstack{{\miniscule (\rule[.5ex]{1.25em}{.18mm})}
  \\ [-.7ex] $#1$}}

\def\uquark    {{\ensuremath{\Pu}}\xspace}

\def\squark    {{\ensuremath{\Ps}}\xspace}

\def\cquark    {{\ensuremath{\Pc}}\xspace}

\def\bquark    {{\ensuremath{\Pb}}\xspace}

\def\pion   {{\ensuremath{\Ppi}}\xspace}

\def\pip    {{\ensuremath{\pion^+}}\xspace}
\def\pim    {{\ensuremath{\pion^-}}\xspace}

\def\kaon    {{\ensuremath{\PK}}\xspace}

\def\KorKbar {\kern \thebaroffset\optbar{\kern -\thebaroffset \PK}{}\xspace}

\def\Kp      {{\ensuremath{\kaon^+}}\xspace}
\def\Km      {{\ensuremath{\kaon^-}}\xspace}

\def\KS      {{\ensuremath{\kaon^0_{\mathrm{S}}}}\xspace}

\def\Kstarp  {{\ensuremath{\kaon^{*+}}}\xspace}

\def\Dbar    {{\ensuremath{\offsetoverline{\PD}}}\xspace}
\def\D       {{\ensuremath{\PD}}\xspace}

\def\DorDbar {\kern \thebaroffset\optbar{\kern -\thebaroffset \PD}\xspace}
\def\Dz      {{\ensuremath{\D^0}}\xspace}
\def\Dzb     {{\ensuremath{\Dbar{}^0}}\xspace}
\def\Dp      {{\ensuremath{\D^+}}\xspace}
\def\Dm      {{\ensuremath{\D^-}}\xspace}

\def\DpDm    {\ensuremath{\Dp {\kern -0.16em \Dm}}\xspace}
\def\Dstar   {{\ensuremath{\D^*}}\xspace}

\def\Dsm     {{\ensuremath{\D^-_\squark}}\xspace}

\def\B       {{\ensuremath{\PB}}\xspace}
\def\Bbar    {{\ensuremath{\offsetoverline{\PB}}}\xspace}

\def\BorBbar {\kern \thebaroffset\optbar{\kern -\thebaroffset \PB}\xspace}
\def\Bz      {{\ensuremath{\B^0}}\xspace}

\def\Bd      {{\ensuremath{\B^0}}\xspace}

\def\BdorBdbar {\kern \thebaroffset\optbar{\kern -\thebaroffset \Bd}\xspace}

\def\Bs      {{\ensuremath{\B^0_\squark}}\xspace}
\def\Bsb     {{\ensuremath{\Bbar{}^0_\squark}}\xspace}
\def\BsorBsbar {\kern \thebaroffset\optbar{\kern -\thebaroffset \Bs}\xspace}

\def\Y#1S{\ensuremath{\PUpsilon{(#1S)}}\xspace}

\def\proton      {{\ensuremath{\Pp}}\xspace}
\def\antiproton  {{\ensuremath{\overline \proton}}\xspace}

\def\Lz          {{\ensuremath{\PLambda}}\xspace}
\def\Lbar        {{\ensuremath{\offsetoverline{\PLambda}}}\xspace}
\def\LorLbar     {\kern \thebaroffset\optbar{\kern -\thebaroffset \PLambda}\xspace}
\def\Lambdares   {{\ensuremath{\PLambda}}\xspace}

\def\Sigmares    {{\ensuremath{\PSigma}}\xspace}

\def\Xires       {{\ensuremath{\PXi}}\xspace}

\def\Xiresbar       {{\ensuremath{\offsetoverline{\Xires}}}\xspace}

\def\Lb           {{\ensuremath{\Lz^0_\bquark}}\xspace}
\def\Lbbar        {{\ensuremath{\Lbar{}^0_\bquark}}\xspace}

\def\Xibm         {{\ensuremath{\Xires^-_\bquark}}\xspace}

\def\Xibbarp      {{\ensuremath{\Xiresbar{}_\bquark^+}}\xspace}

\newcommand{\decay}[2]{\ensuremath{#1\!\to #2}\xspace} 

\def\to                 {\ensuremath{\rightarrow}\xspace}

\def\CP                {{\ensuremath{C\!P}}\xspace}

\def\CPV               {{\ensuremath{C\!PV}}\xspace}

\def\AT#1     {\ensuremath{A_{\mathrm{T}}^{#1}}\xspace}           

\def\C#1      {\ensuremath{\mathcal{C}_{#1}}\xspace}                       
\def\Cp#1     {\ensuremath{\mathcal{C}_{#1}^{'}}\xspace}                    
\def\Ceff#1   {\ensuremath{\mathcal{C}_{#1}^{\mathrm{(eff)}}}\xspace}        
\def\Cpeff#1  {\ensuremath{\mathcal{C}_{#1}^{'\mathrm{(eff)}}}\xspace}       
\def\Ope#1    {\ensuremath{\mathcal{O}_{#1}}\xspace}                       
\def\Opep#1   {\ensuremath{\mathcal{O}_{#1}^{'}}\xspace}                    

\newcommand{\aunit}[1]{\ensuremath{\mathrm{\,#1}}}       

\newcommand{\tev}{\aunit{Te\kern -0.1em V}\xspace}
\newcommand{\gev}{\aunit{Ge\kern -0.1em V}\xspace}
\newcommand{\mev}{\aunit{Me\kern -0.1em V}\xspace}
\newcommand{\kev}{\aunit{ke\kern -0.1em V}\xspace}
\newcommand{\ev}{\aunit{e\kern -0.1em V}\xspace}
 
\newcommand{\mevc}{\ensuremath{\aunit{Me\kern -0.1em V\!/}c}\xspace}
\newcommand{\gevc}{\ensuremath{\aunit{Ge\kern -0.1em V\!/}c}\xspace}
\newcommand{\mevcc}{\ensuremath{\aunit{Me\kern -0.1em V\!/}c^2}\xspace}
\newcommand{\gevcc}{\ensuremath{\aunit{Ge\kern -0.1em V\!/}c^2}\xspace}

\def\fb   {\ensuremath{\aunit{fb}}\xspace}
\def\invfb   {\ensuremath{\fb^{-1}}\xspace}

\def\gsim{{~\raise.15em\hbox{$>$}\kern-.85em
          \lower.35em\hbox{$\sim$}~}\xspace}
\def\lsim{{~\raise.15em\hbox{$<$}\kern-.85em
          \lower.35em\hbox{$\sim$}~}\xspace}

\def\sqs   {\ensuremath{\protect\sqrt{s}}\xspace}

\def\tell1  {TELL1\xspace}
\def\ukl1   {UKL1\xspace}

\newcommand{\Photo}{}

\begin{document}
\vspace*{4cm}
\title{MIXING AND CP VIOLATION IN BEAUTY AND CHARM AT LHCb}

\author{ ANDREA MERLI$^1$ }

\address{$^1$Universit\`a degli Studi di Milano, Milano, Italy}

\maketitle\abstracts{
We review the most precise tests of the CKM model of flavour physics performed by the \lhcb collaboration during the current year. These include world best measurements of both time-dependent and time-independent \CP violation in charm decays, the first observation of a nonzero mass difference of the neutral charm-meson eigenstates, and the most precise determination to date of the mass difference between the \Bs eigenstates. A new simultaneous combination of measurements of the CKM angle $\gamma$ and of charm decays allows to improve the precision on the decay-width difference of the neutral charm-meson eigenstates by a factor of 2 with respect to the combination of charm results only and to reach the best precision on $\gamma$ from a single experiment. The precision of the measurements is mainly limited by statistics, so further improvement is expected in the future.
}

\section{Introduction}

The charge-parity (\CP) symmetry is minimally violated in nature. It is not only relevant for the understanding of a small set of rare weak processes but it actually bears on one of the most intriguing mysteries of cosmology, namely the fact that the universe contains only matter and not antimatter. While the Standard Model (SM) with the Cabibbo-Kobayashi-Maskawa (CKM) mechanism can account for the current experimental results on \CP violation (\CPV), it fails to explain the cosmological matter–antimatter imbalance. Searching for new sources of \CPV is one of the primary goals of flavour physics. In the framework of the SM of particle physics, \CP symmetry between quarks and antiquarks is broken by a single complex phase in the  CKM quark-mixing matrix. A key consistency test of the SM is to verify the unitarity conditions by over-constraining the CKM matrix with various independent measurements. While the magnitudes of the CKM matrix elements can be determined from the decay rates and mixing frequencies of flavor-changing transitions, measurements of \CP asymmetries permit to determine the phases. \lhcb has collected the world largest dataset of charm and beauty hadrons thanks to their large production cross section at the \lhc energy\cite{LHCb-PAPER-2015-041,LHCb-PAPER-2016-031}, giving the opportunity to play a main role in the experimental flavour physics. In the following we present the last updates on the \CPV and mixing measurements performed by \lhcb collaboration.

\section{Search for \CP violation in \decay{\Lb}{[\Kp\pim]_D \proton \Km} towards a $\gamma$ measurement with a baryonic decay}

Studies of beauty-baryon decays to final states involving a single open-charm meson are promising for measurements of \CPV. The presented measurement\cite{LHCb-PAPER-2021-027} reports the results of a study of \decay{\Lb}{[\Kp\pim]_D \proton \Km} where $D$ represents a superposition of \Dz and \Dzb states, with the objectives of observing for the first time the decay and measuring its \CP asymmetry. This decay is of particular interest since its decay amplitude receives contributions from \decay{\bquark}{\cquark} and \decay{\bquark}{\uquark} amplitudes of similar magnitude, given the CKM suppression between the two $D$ decays. The interference between these two amplitudes, which depends upon the CKM angle $\gamma$, is expected to be large\cite{Atwood:1996ci,Atwood:2000ck}, but the different strong phases associated with the various configurations of polarisation states for the \Lb, proton, and intermediate resonances complicate determination of $\gamma$. The analysis is based on proton-proton collision data collected with the \lhcb detector at $\sqs = 7,8$ and 13\tev, corresponding to a total integrated luminosity of 9\invfb. The suppressed \decay{\Lb}{[\Kp\pim]_D \proton \Km} decay is observed for the first time and the \CP asymmetry
\begin{equation}
    A = \frac{\mathcal{B}(\decay{\Lb}{[\Kp\pim]_D \proton \Km})-\mathcal{B}(\decay{\Lbbar}{[\Km\pip]_D \antiproton \Kp})}{\mathcal{B}(\decay{\Lb}{[\Kp\pim]_D \proton \Km})+\mathcal{B}(\decay{\Lbbar}{[\Km\pip]_D \antiproton \Kp})} = 0.12\pm0.09{\rm(stat.)}^{+0.02}_{−0.03}{\rm(syst.)}
\end{equation}
is measured to be compatible with 0. Sensitivity to \CPV requires interference between amplitudes involving intermediate \Dz and \Dzb mesons. This interference is amplified in regions of the phase space involving \decay{\Lb}{DX} contributions, where $X$ labels excited \Lambdares states. Therefore the \CP asymmetry is measured also in restricted phase-space region which involves \decay{\Lb}{DX} decays, where an enhanced sensitivity to $\gamma$ is expected. The result is $A = 0.01 \pm 0.16 {\rm(stat.)}^{+0.03}_{−0.02} {\rm(syst.)}$, still compatible with no \CPV observed.

\section{Precise determination of the $\Bs-\Bsb$ oscillation frequency}

Precise measurements of the oscillations frequency between $\Bs-\Bsb$ are needed to minimise the systematic uncertainty of measurements of the CKM angle $\gamma$ through time-dependent analyses\cite{LHCb-PAPER-2017-047} and, if combined with measurements of its analogue for \Bz mesons, $\Delta m_d$, provide constraints on the unitarity of the CKM matrix. Due to the excellent decay vertex resolution and track momentum resolution, the LHCb detector is ideally suited to resolve the fast $\Bs-\Bsb$  oscillations. To determine if the \Bs meson oscillated into its antiparticle (and viceversa), knowledge of the initially produced flavour eigenstate is required. This is achieved by using a combination of several flavour-tagging algorithms that exploit different features of the b-hadron production process. The \lhcb collaboration has recently measured $\Delta m_s$ using \decay{\Bs}{\Dsm\pip} decays collected during 2015--2018, where the \Dsm meson decays into the $\Kp\Km\pim$ or $\pip\pim\pim$ final states\cite{LHCb-PAPER-2021-005}. The result, in natural units, is $\Delta m_s = 17.7683 \pm 0.0051 \pm 0.0032 \,{\rm ps}^{-1}$. Its relative precision, $3 \times 10^{-4}$, is lower than that of the previous world average by a factor of 2. The combination with previous \lhcb measurements yields $\Delta m_s = 17.7656\pm0.0057  \,{\rm ps}^{-1}$, a crucial legacy measurement of the \lhcb collaboration.

\section{Search for \CP violation in \decay{\Xibm}{p\Km\Km} decays}

The first amplitude analysis of a beauty-baryon decay that accounts for \CPV is performed by the \lhcb collaboration with \decay{\Xibm}{\proton\Km\Km} decays\cite{LHCb-PAPER-2020-017}. The analysed dataset has been collected during 2011-2012 ($\sqs = 7,8\tev$) and 2015-2016 (\sqs=13\tev), corresponding to an integrated luminosity of 3\invfb and 2\invfb, respectively. The fitted yields are $193\pm21$ ($297\pm23$) for the former (latter) dataset. The search for \CPV is conducted through a model-dependent amplitude analysis, using data with high signal purity of candidates selected in the signal region ($m(\Xibm)\pm40\mev$). Assuming \Xibm is produced with a negligible polarisation, the phase space of \decay{\Xibm}{\proton\Km\Km} is described using two Dalitz Plot variables. The amplitude model is developed using helicity formalism for the decay dynamics and the isobar formalism to sum all intermediate resonances. The parameters are obtained with a fit to the observed differential decay density $d\Gamma/d\Omega$ in the phase space. The \CP asymmetry parameters associated with each component $i$ of the model are defined as
\begin{equation}
 A^\CP_i =\frac{\int_\Omega (d\Gamma^\Xibm_i/d\Omega - d\Gamma^\Xibbarp_i/d\Omega) d\Omega}{\int_\Omega (d\Gamma^\Xibm_i/d\Omega + d\Gamma^\Xibbarp_i/d\Omega) d\Omega} 
\end{equation}
where $d\Gamma^\Xibm_i/d\Omega$ ($d\Gamma^\Xibbarp_i/d\Omega$) is the particle \Xibm (antiparticle \Xibbarp) differential decay density of the component $i$. The model includes contributions from various \Lambdares and \Sigmares resonances that are well established as well as non-resonances contributions. The former are modelled as Breit-Wigner while the latter as exponential. Six contributions are found to give a good description of the data: \Sigmares(1385), \Lambdares(1405), \Lambdares(1520), \Lambdares(1670), \Sigmares(1775) and \Sigmares(1915) resonances. The \CP asymmetry $A^\CP_i$ for each contributing component $i$ is measured and no significant evidence of \CPV is observed.

\section{Search for time dependent \CP violation in \decay{\Dz}{\Kp\Km} and \decay{\Dz}{\pip\pim}}

A new measurement of the time-dependent \CPV in \decay{\Dz}{\Kp\Km} and \decay{\Dz}{\pip\pim} is performed by the \lhcb collaboration\cite{LHCb-PAPER-2020-045}. These are the same channels that were used for the 2019 $\Delta A_\CP$ discovery\cite{LHCb-PAPER-2019-006}. The \Dz is required to originate from a \decay{\Dstar}{\Dz\pip} decay so that its tag flavour could be deduced from the charge of the tagging pion. The analysis uses proton-proton collision data collected from 2015 to 2018 at a centre-of-mass energy of 13 TeV, corresponding to an integrated luminosity of 6\invfb. The data sample corresponds to 58 million and 18 million \decay{\Dz}{\Kp\Km} and \decay{\Dz}{\pip\pim} events, respectively. The signal purity is high at $\approx95\%$. The observable is the time-dependent \CP asymmetry which can be parametrised as
\begin{equation}
    A_\CP (t) = \frac{\Gamma(\decay{\Dz}{f})-\Gamma(\decay{\Dzb}{f})}{\Gamma(\decay{\Dz}{f})+\Gamma(\decay{\Dzb}{f})} = a_f^d + \Delta Y_f \frac{t}{\tau_D}+\mathcal{O}(x^2,y^2,xy)
\end{equation}
where $a_f^d$ is the \CP asymmetry in the decay, $\tau_D$ is the \Dz lifetime, and $\Delta Y_f$ is the parameter of interest sensitive to time-dependent \CPV. The measured $A_\CP$ asymmetries are corrected by production and detection time-dependent asymmetries caused by trigger requirements. The nuisance asymmetries are mitigated by reweighting the events and equalizing the \Dz/\Dzb kinematics. The procedure is validated on  the control channel \decay{\Dz}{\Km\pip}. The results are $\Delta Y_{\Kp\Km} = (−2.3\pm1.5\pm0.3)\times10^{−4}$ and $\Delta Y_{\pip\pim} = (−4.0\pm2.8\pm0.4)\times10^{−4}$, where the first uncertainty is statistical and second systematic. $\Delta Y_{\Kp\Km}$ and $\Delta Y_{\pip\pim}$ agree with each other within $0.5\sigma$. This agreement is encouraging as the final state dependence is expected to be negligible. The results are also compatible with no \CPV within $2\sigma$. The precision of the results was improved by a factor of two compared with the previous measurement\cite{LHCb-PAPER-2016-063}. A weighted average between these final states, denoted as $\Delta Y$, and combined with previous \lhcb measurements\cite{LHCb-PAPER-2014-069,LHCb-PAPER-2016-063,LHCb-PAPER-2019-032} is $\Delta Y = (-1.0 \pm 1.1 \pm 0.3) \times 10^{-4}$, where the first uncertainty is statistical and the second systematic. This value is consistent with \CP symmetry and constitutes the world most precise determination of this quantity. 

\section{Observation of the mass difference between neutral charm-meson eigenstates}

The decay width difference between neutral charm mesons, $y\neq 0$, has been established in the past years\cite{BaBar:2012bho,Belle:2015etc,LHCb-PAPER-2018-038}. However, the mass difference, $x\neq 0$, has so far been elusive. The most precise past measurement was reported by \lhcb as $x=(0.27^{+0.17}_{-0.15})\times10^{-3}$~\cite{LHCb-PAPER-2019-001}. The analysis presented here\cite{LHCb-PAPER-2021-009} employs a “bin-flip” method\cite{DiCanto:2018tsd} which is optimized for $x$ sensitivity. Like the previous presented analysis, \Dz are selected from \decay{\Dstar}{\Dz\pip} decays. The analysis uses proton-proton collision data collected by the \lhcb experiment from 2016 to 2018, corresponding to an integrated luminosity of 5.4\invfb. A total of 30.6 million \decay{\Dz}{\KS\pip\pim} are analysed. The analysis exploits the rich, resonant structure of the multi-body final state. Many interfering amplitudes such as $\Dz \! \xrightarrow{DCS} \Kstarp \pim \!\to \KS \pip\pim $ and $\Dz \!\xrightarrow{mix} \Dzb \! \xrightarrow{CF} \Kstarp \pim \!\to \KS \pip\pim$ contribute. The Dalitz plane is divided into upper and lower bins, mirrored across the diagonal. They are defined in such a way that the strong-phase $\Dz-\Dzb$ difference varies minimally across each bin. The strong phases are constrained in the fit using CLEO and BES III inputs. The observables are the ratios of events between upper and lower bins. One of the method’s benefits is that most detector nuisance effects cancel in the ratio. The time-dependent fits to the eight bin ratios are performed. Deviations from a constant value are due to mixing, while differences between ratios of initially tagged $\Dz-\Dzb$ are due to \CPV. The resulting mixing values are $x = (3.98^{+0.56}_{-0.54})\times 10^{-3}$ and $y = (4.6^{+1.5}_{-1.4})\times 10^{-3}$, which is the first non-zero measurement of $x$ with a significance larger than $7\sigma$. The parameters sensitive to \CPV in mixing are measured to be $|q/p| = 0.996\pm0.052$ and $\phi = 0.056^{+0.047}_{-0.051}$, consistent with \CP symmetry.

\section{Measurement of \CP asymmetry in \decay{\Dz}{\KS\KS} decays}

The \decay{\Dz}{\KS\KS} decay is ideal for \CP observation because, at leading order, it receives contributions from
two similarly sized amplitudes and $A_\CP$ in this channel might
be large, up to 1\%\cite{Nierste:2015zra}. The presented analysis\cite{LHCb-PAPER-2020-047} boasts several enhancements over the previous analysis that used a smaller dataset\cite{LHCb-PAPER-2018-012}. The nuisance production and detection asymmetries were removed by a weighting technique exploiting the \decay{\Dz}{\Kp\Km} calibration sample. The data was also split into consistent sub-samples based on \KS daughters tracking type, primary interaction origin and other criteria. The new techniques resulted in a 30\% sensitivity improvement. Paired with the dataset size increasing from 2\invfb to 6\invfb, this led to a significantly more precise measurement. The analysis extracted time-integrated $A_\CP$ from a 3D fit to m(\pip\pim) of both \KS candidates and $\Delta m = m(\KS\KS\pip)-m(\KS\KS)$. The result is $A_\CP(\decay{\Dz}{\KS\KS}) = (-3.1\pm1.2\pm0.4\pm0.2)\%$, where the first uncertainty is statistical, the second systematic and the last one comes from the control sample. While the result is compatible with zero within $2.4\sigma$, it is the highest precision measurement of the parameter to date.

\section{Simultaneous determination of CKM angle $\gamma$ and charm mixing parameters}
Deviations between direct measurements of $\gamma$ and the value derived from global CKM fits, which assume validity of the SM and hence unitarity of the CKM matrix, would be a clear indication of physics beyond the SM. Furthermore, comparisons between the value of $\gamma$ measured using decays of different $B$-meson species provide sensitivity to possible new physics effects at tree level given the different decay topologies involved. The experimental uncertainty on $\gamma$ is larger than that obtained from global CKM fits, $\gamma = (65.6^{+0.9}_{−2.7})^\circ$~\cite{Charles:2015gya} using a frequentist framework, and $\gamma = (65.8 \pm 2.2)^\circ$~\cite{UTfit:2006vpt} with a Bayesian approach. Closing this sensitivity gap is a key physics goal of the \lhcb experiment and the comparison between the direct and indirect determinations of $\gamma$ is an important test of the SM. Results from the charm and beauty sectors, based on data collected with the LHCb detector, are combined for the first time. This method provides an improvement on the precision of the charm mixing parameter $y$ by a factor of two with respect to the current world average. The charm mixing parameters are determined to be $x = (0.400^{+0.052}_{−0.053})\%$ and $y = (0.630^{+0.033}_{−0.030})\%$. The angle $\gamma$ is found to be $\gamma = (65.4^{+3.8}_{−4.2})^\circ$ and it represents the most precise determination of this quantity from a single experiment.

\section{Conclusion}

\lhcb collected the largest sample of beauty and charm decays, which led to new world best measurements of \CP asymmetries and mixing parameters. The precision of the measurements is limited mainly by statistics, so further improvement is expected. Many more intriguing \lhcb results are sure to appear in the future, as there are more interesting Run 2 analyses in the pipeline. Finally, Run 3 will start this year and feature higher luminosity and an upgraded detector and trigger.

\section*{Acknowledgments}

A.M. acknowledges support from the European Research Council (ERC) under the European
Union’s Horizon 2020 research and innovation programme under grant agreement No. 771642 (SELDOM) and INFN Sezione di Milano.

\section*{References}

\bibliography{main,LHCb-PAPER}

\begin{thebibliography}{10}

\bibitem{LHCb-PAPER-2015-041}
R.~Aaij et~al.
\newblock {Measurements of prompt charm production cross-sections in
  \proton\proton collisions at $\sqs = $13\tev}.
\newblock {\em JHEP}, 03:159, 2016.

\bibitem{LHCb-PAPER-2016-031}
R.~Aaij et~al.
\newblock {Measurement of the \bquark-quark production cross-section in 7 and
  13$\tev$ $\proton\proton$ collisions}.
\newblock {\em Phys. Rev. Lett.}, 118:052002, 2017.

\bibitem{LHCb-PAPER-2021-027}
R.~Aaij et~al.
\newblock {Observation of the suppressed $\Lb \to \D \Pp \Km$ decay with $\D
  \to \Kp \pim$ and measurement of its \CP asymmetry}.
\newblock {\em Phys. Rev.}, D104:112008, 2021.

\bibitem{Atwood:1996ci}
David Atwood, Isard Dunietz, and Amarjit Soni.
\newblock {Enhanced CP violation with B ---\ensuremath{>} K D0 (anti-D0) modes
  and extraction of the CKM angle gamma}.
\newblock {\em Phys. Rev. Lett.}, 78:3257--3260, 1997.

\bibitem{Atwood:2000ck}
David Atwood, Isard Dunietz, and Amarjit Soni.
\newblock {Improved methods for observing CP violation in B+- ---\ensuremath{>}
  K D and measuring the CKM phase gamma}.
\newblock {\em Phys. Rev. D}, 63:036005, 2001.

\bibitem{LHCb-PAPER-2017-047}
R.~Aaij et~al.
\newblock {Measurement of \CP asymmetry in \mbox{\decay{\Bs}{D_s^\mp \Kpm}}
  decays}.
\newblock {\em JHEP}, 03:059, 2018.

\bibitem{LHCb-PAPER-2021-005}
R.~Aaij et~al.
\newblock {Precise determination of the $B_{s}^{0} - \bar{B}_{s}^{0}$
  oscillation frequency}.
\newblock 2021.
\newblock {to appear in Nature Physics}.

\bibitem{LHCb-PAPER-2020-017}
R.~Aaij et~al.
\newblock {Search for \CP violation in $\Xibm \to p K^- K^- decays$}.
\newblock {\em Phys. Rev.}, D104:052010, 2021.

\bibitem{LHCb-PAPER-2020-045}
R.~Aaij et~al.
\newblock {Search for time-dependent \CP violation in $\Dz \to \Kp \Km$ and
  $\Dz \to \pip \pim$ decays}.
\newblock {\em Phys. Rev.}, D104:072010, 2021.

\bibitem{LHCb-PAPER-2019-006}
R.~Aaij et~al.
\newblock {Observation of \CP violation in charm decays}.
\newblock {\em Phys. Rev. Lett.}, 122:211803, 2019.

\bibitem{LHCb-PAPER-2016-063}
R.~Aaij et~al.
\newblock {Measurement of the \CP violation parameter $A_\Gamma$ in
  \mbox{\decay{\Dz}{\Kp\Km}} and \mbox{\decay{\Dz}{\pip\pim}} decays}.
\newblock {\em Phys. Rev. Lett.}, 118:261803, 2017.

\bibitem{LHCb-PAPER-2014-069}
R.~Aaij et~al.
\newblock {Measurement of indirect \CP asymmetries in
  \mbox{\decay{\Dz}{\Km\Kp}} and \mbox{\decay{\Dz}{\pim\pip}} decays using
  semileptonic \B decays}.
\newblock {\em JHEP}, 04:043, 2015.

\bibitem{LHCb-PAPER-2019-032}
R.~Aaij et~al.
\newblock {Updated measurement of decay-time-dependent \CP asymmetries in
  \mbox{\decay{\Dz}{\Kp\Km}} and \mbox{\decay{\Dz}{\pip\pim}} decays}.
\newblock {\em Phys. Rev.}, D101:012005, 2020.

\bibitem{BaBar:2012bho}
J.~P. Lees et~al.
\newblock {Measurement of $D^0-\bar{D}^0$ Mixing and CP Violation in Two-Body
  $D^0$ Decays}.
\newblock {\em Phys. Rev. D}, 87(1):012004, 2013.

\bibitem{Belle:2015etc}
Marko Stari\v{c} et~al.
\newblock {Measurement of $D^0 – \bar {D^0}$ mixing and search for CP
  violation in $D^0 \to K^+ K^−, \pi^+ \pi^−$ decays with the full Belle
  data set}.
\newblock {\em Phys. Lett. B}, 753:412--418, 2016.

\bibitem{LHCb-PAPER-2018-038}
R.~Aaij et~al.
\newblock {Measurement of the charm-mixing parameter $y_\CP$}.
\newblock {\em Phys. Rev. Lett.}, 122:011802, 2019.

\bibitem{LHCb-PAPER-2019-001}
R.~Aaij et~al.
\newblock {Measurement of the mass difference between neutral charm-meson
  eigenstates}.
\newblock {\em Phys. Rev. Lett.}, 122:231802, 2019.

\bibitem{LHCb-PAPER-2021-009}
R.~Aaij et~al.
\newblock {Observation of the mass difference between neutral charm-meson
  eigenstates}.
\newblock {\em Phys. Rev. Lett.}, 127:111801, 2021.

\bibitem{DiCanto:2018tsd}
A.~Di~Canto, J.~Garra~Tic\'o, T.~Gershon, N.~Jurik, M.~Martinelli,
  T.~Pila\v{r}, S.~Stahl, and D.~Tonelli.
\newblock {Novel method for measuring charm-mixing parameters using multibody
  decays}.
\newblock {\em Phys. Rev. D}, 99(1):012007, 2019.

\bibitem{Nierste:2015zra}
Ulrich Nierste and Stefan Schacht.
\newblock {CP Violation in $D^0\rightarrow K_SK_S$}.
\newblock {\em Phys. Rev. D}, 92(5):054036, 2015.

\bibitem{LHCb-PAPER-2020-047}
R.~Aaij et~al.
\newblock {Measurement of \CP asymmetry in $D^0 \to K_S^0 K_S^0$ decays}.
\newblock {\em Phys. Rev.}, D104:L031102, 2021.

\bibitem{LHCb-PAPER-2018-012}
R.~Aaij et~al.
\newblock {Measurement of the time-integrated \CP asymmetry in
  \mbox{\decay{\Dz}{\KS \KS}} decays}.
\newblock {\em JHEP}, 11:048, 2018.

\bibitem{Charles:2015gya}
J.~Charles et~al.
\newblock {Current status of the Standard Model CKM fit and constraints on
  $\Delta F=2$ New Physics}.
\newblock {\em Phys. Rev. D}, 91(7):073007, 2015.

\bibitem{UTfit:2006vpt}
M.~Bona et~al.
\newblock {The Unitarity Triangle Fit in the Standard Model and Hadronic
  Parameters from Lattice QCD: A Reappraisal after the Measurements of Delta
  m(s) and BR(B ---\ensuremath{>} tau nu(tau))}.
\newblock {\em JHEP}, 10:081, 2006.

\end{thebibliography}






\end{document}